\documentclass[a4paper,12pt,oneside]{article}
\usepackage[latin1]{inputenc}
\usepackage[T1]{fontenc}
\usepackage[english]{babel}
\usepackage[mathscr]{euscript}
\usepackage{amsmath,amsfonts,latexsym,amssymb,exscale,amsthm}
\title{A quantum group for the Einstein equations}
\author{Giuseppe Iurato\footnote{e-mail: iurato@dmi.unict.it}}
\date{}
\begin{document}\maketitle\begin{abstract}In this paper, we expose the construction of a possible, simple matrix quantum group structure (according to Woronowicz), related to elementary formal aspects of the Einstein field equations of General Relativity, and its possible symmetries.\end{abstract}
Mainly, we present a simple application of the results achieved by M. Dubois-Violette and G. Launer in [1], where is built up a first matrix quantum group structure (in the sense of S.L. Woronowics $-$ see [8], 2.1)
associated to an arbitrary non-degenerate bilinear form.\\\\Precisely, we apply, almost verbatim, these considerations to a generalization of the \it Einstein field equations \rm (1915), in purely covariant form given by\footnote{According to the Robertson-Noonan sign convention
(1968) (see [4]).} (see [6], 1.13.5, and [5], 4.0, 4.1)$$G_{ij}=R_{ij}-\frac{1}{2}Rg_{ij}=-8\pi
GT_{ij},\qquad i,j=0,1,2,3,\leqno(1)$$where $G_{ij}$ is the \it Einstein curvature tensor, \rm $R_{ij}$
is the \it Ricci curvature tensor, $g_{ij}$ \rm is the \it Lorentz metric, $R$ \rm is the \it Ricci scalar, \rm $G$
is the \it gravitational constant, \rm and $T_{ij}$ is the so called \it Hilbert tensor \rm (see [9], Chapter 7) with $T_{ij}=T_{ij}(g_{lh},\partial g_{lh};\psi_k,\partial\psi_k)$ in the presence of a set of physical fields $\psi_k\ \ k=1,...,p$.\\\rm In the geometrized units, it is $G=1$ (see [7]).\\\\We recall that the Einstein field equations (1) may be deduced both by a variational Palatini's argument (see [9]) and, inductively, by the newtonian Poisson's equation  $\Delta\phi=4\pi G\rho$.\\Following the latter way, it is assumed that the field equations for the gravitational field, that we may call \it generalized Einstein (field) equations, \rm should have a general form of the type (see [9], Chapter 4, [10], Cap. II, § 2.1, and [4], Chapter 17, § 17.1) $$G_{ij}(g_{lh},g_{lh,r},g_{lh,rt},...)=k\pi T_{ij},\qquad i,j=0,1,2,3,\leqno(2)$$where $G_{ij}$ is a yet to be determined tensor function of the metric tensor $g_{lh}$ and some of its derivatives, and $k$ is a real constant.\\\\ Many physical reasons (see [10], § 2.1, and [4], § 17.1) restricts the class of the possible functions $G_{ij}$, satisfying (2), to a well-defined tensor, namely the Einstein curvature tensor mentioned above, obtaining the known equations (1).\\\\On the other hand, by earlier Weyl's and Cartan's results culminated in Lovelock's statement (see [16]), if we seek a tensor equation of the form $G_{ij}=T_{ij}$, where the components $A_{ij}$ involve the metric tensor $g_{ij}$ and its first and second derivatives (hence, assuring second-order partial differential equations generalizing the Poisson one), and if $A_{ij}$ have vanishing divergence $A_{ij;j}=0$, then the equation must be of the form $aG_{ij}+bg_{ij}=-8\pi kT_{ij}$, where $a$ and $b$ are constants; Einstein's choice is then $a=1,b=0$ ($b$ is said to be the \it cosmological constant\rm).\\\\At this point, taking into account the geometrical meaning of the Einstein's equations\footnote{These arguments shall be the matter of another paper.} according to [17], it is possible to consider the following \it Einstein bilinear form$$\Omega_{ij}\doteq G_{ij}+8\pi T_{ij},\qquad i,j=0,1,2,3,\leqno(2')$$\rm that, for now, we suppose to be non-degenerate; its zero values are the generalized Einstein equations (2). In $(2')$, we suppose, a priori, $G_{ij}$ to be an arbitrary bilinear form (of $\mathbb{R}^4$), while $T_{ij}$ is the Hilbert tensor.\\\\
In [1] (see also [11], Example 4.62), it is considered a finite family $\{T({\alpha})\}_{\alpha\in\Xi}$ of $(r_{\alpha},s_{\alpha})$-tensors on
$\mathbb{R}^n$ and the group $G$ of the automorphisms of $\mathbb{R}^n$ that preserve $T({\alpha})$ in the
following sense
$$u_{k_1}^{i_1}...u_{k_{r_{\alpha}}}^{i_{r_{\alpha}}}T(\alpha)_{j_1...j_{s_{\alpha}}}^{k_1...k_{r_{\alpha}}}=
u_{j_1}^{k_1}...u_{j_{s_{\alpha}}}^{k_{s_{\alpha}}}T(\alpha)_{k_1...k_{s_{\alpha}}}^{i_1...i_{r_{\alpha}}}
\qquad\forall\alpha\in\Xi,\leqno(3)$$supposing invertible the generic matrix $u=\|u_{j}^i\|\in G$.\\\\ In matrix
quantum group theory (see [2]), one can considers the elements $u_j^i$ as linear coordinate functions on $G$,
which assigns to each $g\in G$ its matrix elements (respect to a given base), namely $u_j^i(g)=g_j^i$, and that one
can also interprets as generating the unital associative algebra $Fun (G)$ of functions on $G$, under the relations (3).\\ The latter is a commutative Hopf algebra, with usual comultiplication given by $(\Delta(f))(g_1,g_2)=f(g_1g_2)$, so
that the cocommutativity, or not, of this algebra, is related to the commutativity, or not, of the group $G$;
furthermore, the coproduct is induced by $\Delta u_j^i=u_k^i\otimes u^k_j$, since $u_j^i(g)=g_j^i$.\\\\Hence,
following [1], we could say that (3) defines a first (matrix) quantum group structure preserving each $T(\alpha)$; moreover,
we restricts our study to the case in which $T(\alpha)$ is a given non-degenerate bilinear form $\Omega_{ij}$ on
$\mathbb{R}^4$, with dual $\Omega^{ij}$ (given by the inverse matrix), that is we suppose $r_{\alpha}=0,s_{\alpha}=2$, $card\ \Xi=1$ and\footnote{However, the following considerations holds true also for any $n\geq 2$.} $n=4$.\\\\If $\Omega$ is a bilinear form on
$\mathbb{R}^4$ with components (respect to a given base) $\Omega_{ij}$, and $\tilde{\Omega}$ is a bilinear form
on its dual with components (respect to the dual base) $\tilde{\Omega}^{ij}$, then, as known,
$\tilde{\Omega}\otimes\Omega$ is identified with the endomorphisms of $\mathbb{R}^4\otimes\mathbb{R}^4$ with
components $\Omega^{i_1i_2}\Omega_{j_1j_2}$; likewise, if $u$ and $v$ are endomorphisms of $\mathbb{R}^4$ with
components $u_j^i$ and $v_j^i$, then $u\otimes v$ is identified with the endomorphism of
$\mathbb{R}^4\otimes\mathbb{R}^4$ with components $u_{j_1}^{i_1}v_{j_2}^{i_2}$.\\\\Let $\Omega$ be the
non-degenerate bilinear form with components (in the canonical base) $\Omega_{ij}$ given by (3); the matrix of
its components $\Omega_{ij}$, will be denoted again by $\Omega$. Associated to $\Omega$ is its dual
$\Omega^{-1}$ of $\mathbb{R}^4\otimes\mathbb{R}^4$, that is the bilinear form on the dual of $\mathbb{R}^4$
($\cong\mathbb{R}^4$), with components $\Omega^{ij}$ defined by $\Omega^{ik}\Omega_{kj}=\delta_j^i$; the matrix
of the components $\Omega^{ij}$ will be again denoted by $\Omega^{-1}$, the inverse of the matrix $\Omega$ (that
there exists because $\Omega$ is non-degenerate).\\\\Let $\mathcal{A}_{\mathbb{R}}(\Omega)$ be the unital associative
$\mathbb{R}$-algebra generated by the scalars $t_j^i\in\mathbb{R}\ \ i,j=0,1,2,3$, with the
relations
$$\Omega_{ij}t_k^it_l^j=\Omega_{kl},\qquad \Omega^{ij}t^k_it_j^l=\Omega^{kl},\qquad k,l=0,1,2,3,$$where
$\Omega_{kl},\Omega^{kl}\in\mathbb{R}$ are identified, respectively, with
$\Omega_{kl}1_{\mathcal{A}},\Omega^{kl}1_{\mathcal{A}}\in\mathcal{A}_{\mathbb{R}}(\Omega)$, if $1_{\mathcal{A}}$
is the unit of $\mathcal{A}_{\mathbb{R}}(\Omega)$.\\Hence, it is possible to prove (see [1])
that\begin{enumerate}\item there exists a unique homomorphism of algebras, say
$\Delta:\mathcal{A}_{\mathbb{R}}(\Omega)\rightarrow\mathcal{A}_{\mathbb{R}}(\Omega)\otimes\mathcal{A}_{\mathbb{R}}
(\Omega)$, such that $\Delta t_j^i=t_k^i\otimes t_j^k\ \ i,j=0,1,2,3$;\item there exists a unique homomorphism
of algebras, say $\varepsilon:\mathcal{A}_{\mathbb{R}}(\Omega)\rightarrow\mathbb{R}$, such that
$\varepsilon(t_j^i)=\delta_j^i\ \ i,j=0,1,2,3$;\item there exists a unique linear antimultiplicative mapping,
say $S:\mathcal{A}_{\mathbb{R}}(\Omega)\rightarrow\mathcal{A}_{\mathbb{R}}(\Omega)$, such that\footnote{Setting $t=\|t_j^i\|$, it is $S(t)=(\Omega^{-1})^{t}t\Omega$.}
$S(t_j^i)=\Omega^{ik}t_k^l\Omega_{lj}\ \ i,j=0,1,2,3,$ and $S(1_{\mathcal{A}})=1_{\mathcal{A}}$.
\end{enumerate}Furthermore, $\Delta$ is a coproduct, $\varepsilon$ is a counit, and $S$ is an antipode\footnote{In general, there is no antipode for a generic tensor $T(\alpha)$.} since
$S(t_k^i)t_j^k=t^i_kS(t_j^k)$, so that, denoted by $m$ the product of the algebra
$\mathcal{A}_{\mathbb{R}}(\Omega)$, we have that
$(\mathcal{A}_{\mathbb{R}}(\Omega),m,1_{\mathcal{A}},\Delta,\varepsilon,S)$ is a Hopf algebra, called the \it Hopf algebra of the Einstein bilinear form $\Omega$.\\\\ \rm This Hopf algebra defines (in the terminology of
[1]; see also [12], Appendix 2) the quantum group of the non-degenerate bilinear form $\Omega$, that we may call the \it Einstein quantum group\rm; hence, as usual, we may think $\mathcal{A}_{\mathbb{R}}(\Omega)$ as a kind of algebra of 'representative functions' on this quantum group.\\\\Besides, this quantum group extends the classical group of the linear transformations of
$\mathbb{R}^4$ which preserves $\Omega$, and, therefore, such a quantum object may represents further generalized
symmetries of the Einstein bilinear form $\Omega$.\\\\The matrix $t=\|t_j^i\|\in
M^{(4,4)}(\mathcal{A}_{\mathbb{R}}(\Omega))$ is a multiplicative matrix (see [3]) whose entries generates
$\mathcal{A}_{\mathbb{R}}(\Omega)$, obtaining an example of matrix quantum group.\\\\\it Note. \rm All the above considerations about $\mathcal{A}_{\mathbb{R}}(\Omega)$, holds for an arbitrary non-degenerate bilinear form $\Omega$ of $\mathbb{R}^n$, with $n\geq 2$.\\\\Given a non-degenerate bilinear form $\Omega$ on $\mathbb{R}^n$ ($n\geq 2$), with components $\Omega_{ij}$ (respect to the canonical base), we may define the quadratic homogeneous algebra\footnote{For brief recalls on homogeneous algebras, see [13] or [12], Appendix 1, and references therein.} $\mathcal{Q}_{\mathbb{R}}(\Omega)$ generated by the elements $x^j\ \ j=1,...,n$, with the relations $\Omega_{ij}x^ix^j=0$.\\In [12], § 2. (see also [13]), it is proved as $\mathcal{Q}_{\mathbb{R}}(\Omega)$ be a Gorenstein and Koszul algebra of global dimension 2. Conversely, it is possible to prove that any quadratic algebra generated by $n$ elements $x^j$, finitely generated in degree 1 and finitely presented with relations of degree $\geq 2$, which is Gorenstein and Koszul of low global dimension 2, is an algebra of the type $\mathcal{Q}_{\mathbb{R}}(\Omega)$ for a certain non-degenerate bilinear form $\Omega$.\\\\Moreover, if $\Omega\stackrel{\chi}{\rightarrow}\Omega\circ M$ is the action given by $(\Omega\circ M)(x,y)=\Omega(Mx,My)$ for each $M\in GL_n(\mathbb{R})$ and $x,y\in\mathbb{R}^n$, then it follows that $\chi$ preserves the non-degeneracy of bilinear forms, and $\mathcal{Q}_{\mathbb{R}}(\Omega)\cong\mathcal{Q}_{\mathbb{R}}(\Omega')$ if and only if $\Omega$ and $\Omega'$ belong to the same $GL_n(\mathbb{R})$-orbit of $\chi$, that is, if and only if $\Omega'=\Omega\circ M$ for some $M\in GL_n(\mathbb{R})$.\\Therefore, since the action of $\chi$ corresponds to a change of generators in $\mathcal{A}_{\mathbb{R}}(\Omega)$, it follows that $\mathcal{A}_{\mathbb{R}}(\Omega)$ only depends by the orbit of $\Omega$ under $\chi$.\\So, we may define the \it moduli space $\mathcal{M}(\mathcal{Q}_{\mathbb{R}}(\Omega))$ \rm of $\mathcal{Q}_{\mathbb{R}}(\Omega)$, to be the space of all $GL_n(\mathbb{R})$-orbits of $\chi$.\\\\Furthermore, taking into account what has been said above about $\mathcal{A}_{\mathbb{R}}(\Omega)$ in $\mathbb{R}^n$, by Proposition 20 of [12], Appendix 2, follows that there is a unique algebra homomorphism $\Delta_t:\mathcal{Q}_{\mathbb{R}}(\Omega)\rightarrow\mathcal{A}_{\mathbb{R}}(\Omega)\otimes
\mathcal{Q}_{\mathbb{R}}(\Omega)$ such that $\Delta_t(x^j)=t^j_i\otimes x^i$ for all $j=1,...,n$, endowing $\mathcal{Q}_{\mathbb{R}}(\Omega)$ of a $\mathcal{A}_{\mathbb{R}}(\Omega)$-comodule structure.\\Hence, the quantum group of $\Omega$ coacts on the quantum space corresponding to $\mathcal{Q}_{\mathbb{R}}(\Omega)$, that is $\mathcal{Q}_{\mathbb{R}}(\Omega)$ corresponds to the natural quantum space for the coaction of $\mathcal{A}_{\mathbb{R}}(\Omega)$.\\\\Come back to the case $n=4$, in [1] the Hopf algebra $\mathcal{A}_{\mathbb{R}}(\Omega)$ is also endowed with a particular quasi-triangular structure through a $R$-matrix, say
$\mathcal{R}:\mathbb{R}^4\otimes\mathbb{R}^4\rightarrow\mathbb{R}^4\otimes\mathbb{R}^4$, given by
$\mathcal{R}_a=\tau+a(\Omega^{-1})^t\otimes\Omega$, where $a\in\mathbb{R}\setminus\{0\}$ and $\tau$ is the flip
map.\\\\Indeed, for $a\neq 0$, we have the following homogeneous defining relations of
$\mathcal{A}_{\mathbb{R}}(\Omega)$:
$$\mathcal{R}^{i_1i_2}_{k_1k_2}t^{k_1}_{j_1}t^{k_2}_{j_2}=t^{i_1}_{k_1}t^{i_2}_{k_2}\mathcal{R}^{k_1k_2}_{j_1j_2},
\qquad i_l,j_l=0,1,2,3,\quad l=1,2,$$ so that such $\mathcal{R}$ is a $R$-matrix because it satisfy the
following, well-known Yang-Baxter equation
$\mathcal{R}_{12}\mathcal{R}_{13}\mathcal{R}_{23}=\mathcal{R}_{23}\mathcal{R}_{13}\mathcal{R}_{12}$ when and
only when $a\in\mathbb{R}\setminus\{0\}$ verify the braid relation $a(1+a\Omega^{ij}\Omega_{ij}+a^2)=0$, equivalent (since $a\neq 0$) to $a+a^{-1}+\Omega^{ij}\Omega_{ij}=a+a^{-1}+tr(\Omega^{-1}\Omega^t)=0$.\\Thus, we have a $R$-matrix for $\mathcal{A}_{\mathbb{R}}(\Omega)$, given by $\mathcal{R}=\tau+a(\Omega^{-1})^t\otimes\Omega$
with $a\in\mathbb{R}\setminus\{0\}$ such that $a+a^{-1}+\Omega^{ij}\Omega_{ij}=0$.\\\\In [14], it is proved that the representation category of $\mathcal{A}_{\mathbb{R}}(\Omega)$ (in $\mathbb{R}^n,\ \ n\geq 2$) is monoidally equivalent to the representation category of the quantum group $\mathcal{O}_a(SL_2(\mathbb{R}))$ of functions over $SL_2(\mathbb{R})$, if $a\in\mathbb{R}\setminus\{0\}$ verify the above braid relation, so that $Comod(\mathcal{A}_{\mathbb{R}}(\Omega))\cong^{\otimes}Comod(O_q(SL_2(\mathbb{R}))$.\\Moreover, in the § 5. of [14] it is also presented the following isomorphic classification of the Hopf algebra $\mathcal{A}_{\mathbb{R}}(\Omega)$: if $\Omega$ and $\Omega'$ are non-degenerate bilinear forms respectively in $\mathbb{R}^n$ and $\mathbb{R}^m$ with $ n,m\geq 2$, then $\mathcal{A}_{\mathbb{R}}(\Omega)$ and $\mathcal{A}_{\mathbb{R}}(\Omega')$ are isomorphic if and only if $m=n$ and there exists $M\in GL_n(\mathbb{R})$ such that $\Omega'=M^t\Omega M$.\\Then, in the § 6 of [14],
the Author determines the possible Hopf $\ast$-algebra structures and CQG (compact quantum group) algebra structures on $\mathcal{A}_{\mathbb{C}}(\Omega)$ (that is, in the complex case).\\Following the results of [14], T. Aubriot, in [15], studies the possible Galois and bi-Galois objects over $\mathcal{A}_{\mathbb{R}}(\Omega)$.\\\\At last, the paper [1] finishes with some remarks; in particular, the Authors notices that, in dimension $n\geq 3$, there is no $\Omega$ such that $\mathcal{A}_{\mathbb{R}}(\Omega)$ be commutative, that is to say, a such Hopf algebra is necessarily non-commutative.\\\\On the other hand, we remember that in $\mathbb{R}^4$ may be establish a standard canonical complex structure as follows.\\Respect to the canonical base of $\mathbb{R}^4$, if $J_0\in End\ (\mathbb{R}^4)$ is defined putting $J_0(e_j)=e_{n+j}$ for $1\leq j\leq 2$ and $J_0(e_j)=-e_{j-n}$ for $3\leq j\leq 4$, then it follows that such a $J_0$ is a complex structure\footnote{Since $J_0^2=-id_{\mathbb{R}^4}$.} on $\mathbb{R}^4$, and if $\mathbb{R}^4_{\mathbb{C}}(J_0)$ is the resulting
linear complex space structure induced by $J_0$ on $\mathbb{R}^4$, then we have the canonical isomorphism $\mathbb{R}^4_{\mathbb{C}}(J_0)\cong\mathbb{C}^2$.\\From here, it is possible to construct the following faithful representation $\rho:M^{(2,2)}(\mathbb{C})\rightarrow M^{(4,4)}(\mathbb{R})$ defined by$$\rho(A+iB)=\left(\begin{array}{cc}A&-B\\B&A\end{array}\right),$$that it is a $\mathbb{R}$-algebra monomorphism such that $\rho(iH)=J_0\rho(H)$ for any $H\in M^{(2,2)}(\mathbb{C})$, extending the usual immersion\footnote{For any $n\geq 2$, we remember that there exists a well-known immersion $GL_n(\mathbb{C})\hookrightarrow GL_{2n}(\mathbb{R})$.} $GL_2(\mathbb{C})\hookrightarrow GL_{4}(\mathbb{R})$.\\\\Hence, if $\Omega_{ij}\in M^{(4,4)}(\mathbb{R})$ of $(2')$, is such that $\Omega_{ij}\in\rho(M^{(2,2)}(\mathbb{C}))$, let $\tilde{\Omega}_{ij}=\rho^{-1}(\Omega_{ij})\in M^{(2,2)}(\mathbb{C})$; whence, we may identifies $\Omega_{ij}$ with $\tilde{\Omega}_{ij}$, that it is a non-degenerate (if such is $\Omega_{ij}$) bilinear form of $\mathbb{C}^2$.\\Therefore, if $\mathcal{A}_{\mathbb{C}}(\tilde{\Omega})$ is the Hopf algebra associated to $\tilde{\Omega}$, then it is immediate to prove that $\mathcal{A}_{\mathbb{R}}(\Omega)\cong\mathcal{A}_{\mathbb{C}}(\tilde{\Omega})$.\\ In [1], § 6., there is a complete classification of the moduli space of $\tilde{\Omega}$, according to the rank of $\tilde{\Omega}$. Precisely\begin{itemize}\item if $rk\ \tilde{\Omega}=0$, then there is only one orbit of which one representative element is $\left(\begin{array}{cc}0&-1\\1&0\end{array}\right)$, this case corresponding to $SL_2(\mathbb{C})$ with $R$-matrix the identity $R_0$ of $\mathbb{C}^2\otimes\mathbb{C}^2$;\item if $rk\ \tilde{\Omega}=1$, then there is only one orbit of which one representative element is $\left(\begin{array}{cc}0&-1\\1&\lambda\end{array}\right)$ with $\lambda\neq 0$ (these are all equivalent among them), this case corresponding to the so called \it Manin's jordanian \rm (that it is a special quantum deformation of $SL_2(\mathbb{C})$; see [3]), with equivalent $R$-matrices $R_{\lambda}$ such that $\lim_{\lambda\rightarrow 0}R_{\lambda}=R_0$;\item if $rk\ \tilde{\Omega}=2$, then there are many orbits, each represented by $\tilde{\Omega}_q=\left(\begin{array}{cc}0&-1\\q&0\end{array}\right)$ for every $q\in\mathbb{C}\setminus\{0,1\}$, with $\mathcal{A}_{\mathbb{C}}(\tilde{\Omega}_q)\cong SL_{2,q}(\mathbb{C})$, $R$-matrix corresponding to that of $M_{2,q}(\mathbb{C})$ (quantum deformation of $M^{(2,2)}(\mathbb{C})$; see [3]), and $\mathcal{Q}_{\mathbb{C}}(\tilde{\Omega}_q)$ corresponding to the \it Manin plane \rm (that it is the natural quantum space for the coaction of $SL_{2,q}(\mathbb{C})$).\end{itemize}The considerations of this paper, may have physical interpretations in view of the possible physical meaning of $\Omega$ (and of $\tilde{\Omega}$, when $\tilde{\Omega}$ exists).$$\bf References.$$\rm{[1]} M. Dubois-Violette, G. Launer, ''The quantum group of a non-degenerate bilinear form'', Physics Letters B, 245(2) (1990) 175-177.\\\\{[2]} S. Majid, Foundations of quantum group theory, Cambridge University Press, Cambridge, 1995.\\\\{[3]} Yu.I. Manin, Quantum groups and Non-Commutative Geometry, Publications du CRM de
l'Univesité de Montréal, Montréal, 1988.\\\\{[4]} C.W. Misner, K.S. Thorne, J.A. Wheeler, Gravitation,
W.H. Freeman and Company, San Francisco, 1973.\\\\{[5]} R.K. Sachs, H. Wu, General Relativity for
Mathematicians, Springer-Verlag, New York, 1977.\\\\{[6]} J. Stewart, Advanced General Relativity,
Cambridge University Press, Cambridge, 1991.\\\\{[7]} R.M. Wald, General Relativity, University of Chicago
Press, Chicago, 1984.\\\\{[8]} M. Dubois-Violette, ''On the theory of quantum groups'', Letters in Mathematical Physics, 19 (1990) 121-126.\\\\{[9]} M. Francaviglia, Relativistic Theories, Quaderni del GNFM-CNR, Firenze, 1988.\\\\{[10]} L. Nobili, Astrofisica Relativistica, CLEUP Editrice, Padova, 2003.\\\\{[11]} J. Madore, An Introduction to Noncommutative Geometry and its Physical Applications, LMS 206, Cambridge University Press, Cambridge, 1998.\\\\{[12]} M. Dubois-Violette, ''Multilinear Forms and Graded Algebras'', Journal of Algebra, 317 (2007) 198-225.\\\\{[13]} M. Dubois-Violette, ''Graded algebras and multilinear forms'', C.R. Acad. Sci. Paris, Ser. I, 341 (2005) 719-724.\\\\{[14]}
J. Bichon, ''The representation category of the quantum group of a non-degenerate bilinear form'', Comm. Alg., 31 (2003) 4831-4851.\\\\{[15]} T. Aubriot, ''On the classification of Galois objects over the quantum group of a nondegenerate bilinear form'', Man. Math., 122 (2007) 119-135.\\\\{[16]} D. Lovelock, ''The four-dimensionality of space and the Einstein's tensor'', Journal of Mathematical Physics, 13 (6) (1972) 874-876.\\\\{[17]} T. Frankel, Gravitational Curvature, W.H. Freeman and Comp., San Francisco, 1979.

\end{document}